\documentclass[conference]{IEEEtran}
\IEEEoverridecommandlockouts
\usepackage{amsmath,amssymb,amsfonts}
\usepackage{algorithmic}
\usepackage{graphicx}
\usepackage{textcomp}
\usepackage{xcolor}

\usepackage[normalem]{ulem}
\usepackage{inconsolata}

\usepackage{tkz-graph}    
    
\usepackage{pgfplots}

\usepackage{tikz}
\usetikzlibrary{patterns}
\usetikzlibrary{arrows}
\usetikzlibrary{calc}
\usetikzlibrary{shapes}

\usepackage{dblfloatfix}
\usepackage{pgfkeys}
\usetikzlibrary{backgrounds}
\pgfplotsset{compat=newest}
\usepackage[export]{adjustbox}
    
\usepackage{comment}
    
\usepackage{xspace}

\usepackage{algorithm}
\usepackage{algorithmic,multicol}

\usepackage{multirow}
\usepackage{booktabs}

\usepackage{caption} 
\usepackage{authblk}

\usepackage[backend=biber,
sorting=none,
doi=false,
isbn=false,
url=false,
maxnames=2,
minnames=2,
style=ieee]{biblatex}
\newcommand{\toolorig}{\textsc{SeeDot}\xspace}
\newcommand{\tool}{\textsc{MAFIA}\xspace}
\newcommand{\protonn}{\textsc{ProtoNN}\xspace}
\newcommand{\bonsai}{\textsc{Bonsai}\xspace}

\newcommand{\vivado}{\textsf{Vivado No Opt}\xspace}
\newcommand{\vivadoauto}{\textsf{Vivado Auto Opt}\xspace}

\newcommand{\vivadoplus}{\textsf{Vivado + \tool}\xspace}
\newcommand{\mafia}{\textsf{\tool}\xspace}

\definecolor{bblue}{HTML}{4F81BD}
\definecolor{rred}{HTML}{C0504D}
\definecolor{ggreen}{HTML}{9BBB59}
\definecolor{ppurple}{HTML}{9F4C7C}
\definecolor{yyellow}{HTML}{DDAE06}

\tikzstyle{bar1} = [bblue, fill=bblue, postaction={pattern=north east lines}]
\tikzstyle{bar2} = [rred, fill=rred]
\tikzstyle{bar3} = [ggreen, fill=ggreen, postaction={pattern=north west lines}]
\tikzstyle{bar4} = [ppurple, fill=ppurple, postaction={pattern=crosshatch dots}]
\tikzstyle{bar5} = [yyellow, fill=yyellow, postaction={pattern=grid}]

\begin{document}

\title{MAFIA: Machine Learning Acceleration on FPGAs for IoT Applications
}


\author[]{Nikhil P Ghanathe\textsuperscript{1}}

\author[]{Vivek Seshadri\textsuperscript{2}}

\author[]{Rahul Sharma\textsuperscript{2}}

\author[]{Steve Wilton\textsuperscript{1}}

\author[]{Aayan Kumar\textsuperscript{3}}

\affil[]{\textsuperscript{1}\textit{University of British Columbia}\quad \textsuperscript{2}\textit{Microsoft Research}\quad \textsuperscript{3}\textit{University of California, Berkeley}}








\maketitle


\begin{abstract}

Recent breakthroughs in ML have produced new classes of models that allow ML inference to run directly on milliwatt-powered IoT devices. On one hand, existing ML-to-FPGA compilers are designed for deep neural-networks on large FPGAs. On the other hand, general-purpose HLS tools fail to exploit properties specific to ML inference, thereby resulting in suboptimal performance. We propose \tool, a tool to compile ML inference on small form-factor FPGAs for IoT applications. \tool provides native support for linear  algebra  operations and can express a variety of ML algorithms, including state-of-the-art models. We show that \tool-generated programs outperform best-performing variant of a commercial HLS compiler by 2.5$\times$ on average.

\end{abstract}

\section{Introduction}
\label{sec:introduction}

Traditionally, IoT devices are used for data collection and the data analysis is performed in the cloud~\cite{cloud1, cloud2, cloud3}. However, performing ML inference directly on IoT devices has recently gained attention, offering benefits such as real-time analysis, increased privacy, and reduced energy consumption. Recent breakthroughs have produced new classes of ML applications that have compute and memory requirements low enough to be run directly on milliwatt-scale IoT devices~\cite{protonn,bonsai} while still having high classification accuracy. While microcontrollers are often used to implement these applications~\cite{gesturepod,farmbeats}, reconfigurable devices like FPGAs may be better suited for such applications. Several reconfigurable architectures for edge computing have been proposed by prior works~\cite{reconfig-on-edge-1, reconfig-on-edge-2,  reconfig-on-edge-3}. On one hand, as we can configure an FPGA to directly run a program, it can deliver superior performance and energy efficiency compared to general-purpose micro-controllers. On the other hand, unlike ASICs which do not allow updates to algorithms once deployed and incur a high non-recurring engineering (NRE) cost, an FPGA can be reprogrammed in the field using over-the-air updates~\cite{ota_1,ota_2} to accommodate the constant evolution of ML models. Unfortunately, FPGAs are notoriously hard to program, even for experts. 

To enable practical use of FPGAs in this domain, we need a compiler that can take high-level specifications of a ML model and directly compile it to an FPGA program. Unfortunately, existing solutions are insufficient in that they 
mainly focus on running DNN models on large FPGAs. However, state-of-the-art ML models~\cite{protonn,bonsai} for IoT applications are mostly based on classical ML algorithms~\cite{mitchell}, and not DNNs. In fact, existing DNN-to-FPGA compilers cannot even express the state-of-the-art ML models designed for IoT applications, and the techniques used by DNN-to-FPGA compilers do not extend to small form-factor FPGAs because they assume an abundance of FPGA resources that allows them to preconfigure the FPGA as, for instance, a sea of matrix multipliers. While general-purpose C-HLS tools can express our target models, they do not exploit specific properties of ML inference and thereby generate suboptimal programs. Section~\ref{sec:motivation} describes the limitations of these two approaches in more detail.

We propose \tool, a compiler for Machine-learning Acceleration on FPGA for IoT Applications. \tool provides native support for linear algebra operations and can express classical ML inference algorithms~\cite{mitchell} like decision trees~\cite{decisiontree} and k-nearest neighbors~\cite{kNN}.
We currently support a subset of  TensorFlow~\cite{tensorflow} and the framework of a recent prior work, \toolorig~\cite{Seedot} that compiles ML inference code to IoT devices (Section~\ref{sec:main-artifacts}).
We compare \tool with four different tools: A)~Bambu HLS, B)~Xilinx Vivado HLS, C)~Vivado HLS with automatically generated compiler hints, and D)~(C) with additional manual hints. We evaluate these mechanisms on two state-of-the-art ML models~\cite{protonn,bonsai}, each on ten different data sets used as benchmark by prior works~\cite{bonsai,protonn,Seedot}. \tool-generated programs consistently outperform all prior approaches -- 2.5$\times$ better on average compared to Vivado HLS with automatically generated hints.

\section{Limitations of Existing HLS Tools}
\label{sec:motivation}

\subsection{ML-HLS tools target the Neural-network family}
\label{sec:limit-mlhls}

Existing ML-HLS tools such as DNN Weaver~\cite{dnnweaver}, FP-DNN~\cite{fp-dnn} and DeepBurning~\cite{deepburning} specifically target deep neural network (DNN) workloads.  There are several reasons these tools do not work well for our target models.

    \vspace{1mm}\noindent \textbf{Building Blocks}: These tools focus on DNN layers (e.g. conv, pooling, FC). As a result, currently, they cannot  express classical algorithms such as decision tree, k-Nearest Neighbors, and kernels like RBF of SVM. Although, we can incorporate support for the classical algorithms in these frameworks, the optimization strategies for DNN's are at odds with classical algorithms (see below). \tool identifies that matrix operations are the fundamental building blocks of most classical ML algorithms and expresses the input program as a matrix data flow graph (DFG).
    
    \vspace{1mm}\noindent \textbf{Parallelism}: The logical view of a DNN model is serial. 
    Therefore, ML-HLS tools only exploit intra-layer parallelism since opportunities for inter-layer parallelization are often scarce. \tool exploits both intra-node and inter-node parallelism (Section~\ref{sec:template-library}).
    
    \vspace{1mm}\noindent \textbf{Nature of Computation}: Most DNN workloads are computation-bound. Therefore, the overhead of data shuffling across layers has minimal impact on latency~\cite{reuse}. However, classical ML algorithms are memory-bound, and data shuffling can significantly increase latency. \tool manages this by introducing intelligent design constraints (Section~\ref{sec:inter-node-communication}).
    
    \vspace{1mm}\noindent \textbf{Resource-efficiency}:~Most ML-HLS tools focus primarily on optimizing matrix multiplication. However, in ML models for IoT applications, the dominant kernel can be any node in the matrix DFG. For example, in the output code of \tool, a matrix addition node may be more critical for latency than a matrix multiplication node. \tool iteratively optimizes the design based on the criticality of each operation in the program (Section~\ref{sec:best-pf-estimator}). Furthermore, existing tools usually target high-end FPGAs and assume abundance of resources. As a result, the optimization techniques employed in these tools are impractical for resource-scarce devices.


\subsection{Limited Scope for Optimizations of C-based HLS tools}
\label{sec:limit-chls}
    \noindent \textbf{Difficult to extract parallelism}:~Automatically extracting parallelism from a sequential C program is difficult. Most safe compiler transformations make conservative assumptions and yield suboptimal performance. 
    To extract better performance, some HLS tools allow the programmer to annotate the C program with compiler hints. However, this requires expert knowledge, is restrictive in the kind of information that can be provided and is time-consuming.
    
\vspace{1mm}\noindent \textbf{Difficult to explore solution space}:~One approach explored by prior work is to automatically generate loop unrolling hints for the HLS compiler. To determine the best unrolling factor for each loop, the compiler must estimate how critical each loop is for the program and the impact of unrolling on resource consumption. This is hard because 1)~finding the critical path in a C program requires knowledge of target domain 2)~HLS compilers have poor accuracy in predicting (without requiring synthesis and simulation) the resource consumption and critical path latency (Section~\ref{sec:regression-model-accuracy}).

\section{Overview of \tool}
\label{sec:overview}

We propose \tool, a compiler that generates high-performance Verilog program from a high-level specification of an ML inference algorithm for a milliwatt-scale FPGA.
The heart of the compiler is an optimizer which minimizes the latency of the critical path in the DFG by
\emph{determining the level of parallelism with which each node in the program's DFG should be executed}.  We formulate this optimization problem as an integer programming problem (Section~\ref{sec:best-pf-estimator}).  





We augment \tool with two components. The first component is a library of Verilog templates with one template for each type of matrix operation. Each template is Verilog implementation of the operation, parameterized by 
the dimensions of the input matrices and a \emph{parallelism factor} (PF). Our Matrix Template Library currently supports various dense and sparse matrix operations: sparse matrix-vector multiplication (SpMV), matrix addition/subtraction, dot product, outer product, hadamard product, dense matrix-vector/vector-matrix/matrix-matrix multiplication, scalar-matrix multiplication and non-linear activations including exponentiation, relu, sigmoid and hyperbolic-tan. These operations are sufficient to express a variety of ML models targeting IoT applications~\cite{edgeml}, including state-of-the-art models~\cite{protonn,bonsai}. \tool's optimizer is agnostic to the exact implementation of the template. Therefore, we can improve performance further by providing an improved template for a particular operation or extend support to other matrix operations by simply adding to the library. 
The second component is a set of latency/resource estimation models that can predict the amount of FPGA resources and execution latency for an operation given its input dimensions and PF. Since there are finite number of operation types, generating these templates and their regression models is tractable and a one-time-effort during the tool development.

\subsection{\tool Compiler Flow}
\label{sec:main-artifacts}

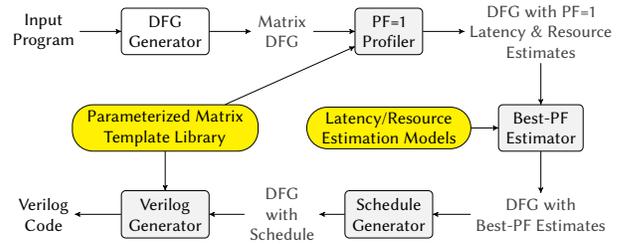
\begin{figure}[tb]
    \centering
    \scalebox{0.55}{\begin{tikzpicture}[font=\sffamily\small\fontfamily{LinuxBiolinumT-LF}\selectfont\large,>=stealth']

  \tikzset{ionode/.style={align=center,minimum width=1.5cm,black!75}};
  \tikzset{procnode/.style={align=center,draw,rounded corners=3pt,inner sep=5pt,fill=black!5}};
  \tikzset{mafialib/.style={align=center,draw,minimum height=1cm,rounded corners=5mm,inner xsep=10pt,fill=yellow}};

  \node (input) [ionode,black] {Input\\Program};
  \node (dfg-gen) at (input.east) [procnode,anchor=west,xshift=1cm,fill=white] {DFG\\Generator};
  \node (dfg) at (dfg-gen.east) [ionode,anchor=west,xshift=1cm] {Matrix\\DFG};
  \node (pf1-prof) at (dfg.east) [procnode,anchor=west,xshift=1cm] {PF=1\\Profiler};
  \node (dfg-pf1) at (pf1-prof.east) [ionode,anchor=west,xshift=1cm] {DFG with PF=1\\Latency \& Resource\\Estimates};
  \node (pf-estimator) at (dfg-pf1.south) [procnode,anchor=north,yshift=-1cm] {Best-PF\\Estimator};
  \node (dfg-best-pf) at (pf-estimator.south) [ionode,anchor=north,yshift=-1cm] {DFG with\\Best-PF Estimates};
  \node (scheduler) at (pf1-prof|-dfg-best-pf) [procnode] {Schedule\\Generator};
  \node (dfg-schedule) at (dfg|-dfg-best-pf) [ionode] {DFG\\with\\Schedule};
  \node (synthesizer) at (dfg-gen|-dfg-best-pf) [procnode] {Verilog\\Generator};
  \node (verilog) at (input|-dfg-best-pf) [ionode,black] {Verilog\\Code};

  \node (ml-model) at (pf1-prof|-pf-estimator) [mafialib] {Latency/Resource\\Estimation Models};
  \node (templates) at (synthesizer|-pf-estimator) [mafialib] {Parameterized Matrix\\Template Library};

  \draw [->] (input) -- (dfg-gen);
  \draw [->] (dfg-gen) -- (dfg);
  \draw [->] (dfg) -- (pf1-prof);
  \draw [->] (pf1-prof) -- (dfg-pf1);
  \draw [->] (dfg-pf1) -- (pf-estimator);
  \draw [->] (pf-estimator) -- (dfg-best-pf);
  \draw [->] (dfg-best-pf) -- (scheduler);
  \draw [->] (scheduler) -- (dfg-schedule);
  \draw [->] (dfg-schedule) -- (synthesizer);
  \draw [->] (synthesizer) -- (verilog);

  \draw [->] (ml-model) -- (pf-estimator);
  \draw [->] (templates.north east) ++(-8mm,0) -- (pf1-prof);
  \draw [->] (templates) -- (synthesizer);

\end{tikzpicture}}
    \caption{Overview of MAFIA Flow}
    \label{fig:mafia-flow}
\end{figure}

Figure~\ref{fig:mafia-flow} shows the compiler's flow.  We provide an overview in this section; more details are provided in Section~\ref{sec:design}.

Given an input program, the \emph{DFG generator} extracts the matrix DFG of the computation along with the matrix dimensions. The \emph{PF-1 Profiler} profiles the resource consumption and latency of each node in the DFG when their PF is set to 1. This information is used by the estimation models to estimate the resource consumption and latency of each node. The \emph{Best-PF Estimator} then uses our optimizer to find the best PF for each node that minimizes the overall latency of the program. 
The \emph{Scheduler Generator} uses the static DFG to generate a schedule that executes the program in \emph{data flow order}. This allows \tool to execute data-independent nodes in parallel. Finally, the \emph{Verilog Generator} uses the identified PFs and the template library to generate the final Verilog code.

\section{Detailed Design}
\label{sec:design}

In this section, we describe \tool in detail. We begin by describing the \emph{Paramterized Matrix Template Library} and the \emph{Latency/Resource Estimation Models} (yellow blocks in Figure~\ref{fig:mafia-flow}). We then explain the flow of the \tool compiler.

\subsection{Parameterized Matrix Template Library}
\label{sec:template-library}

At the backend of the \tool compiler is a library of hand-optimized Verilog templates, one for each type of matrix operation. 
Each template consists of two components: 1)~the execution unit and 2)~the data interface unit.  The execution unit is the core computation unit (eg. adder, multiply-accumulate) with a configurable number of processing elements (PE) specified by the PF.   The data interface unit holds logic for supplying data to the execution unit by reading the output memories of the node's predecessors and writing the result generated by the execution unit into the input memories of the node's consumers. The structure of the data interface unit depends both on the execution PF of the node and the PF of the consumer nodes as follows.

\label{sec:execution-unit}


\label{sec:data-interface-unit}



\label{sec:inter-node-communication}

Consider two nodes: producer and consumer. 
In the producer node, each PE writes its partial output to a buffer. 
However, if the execution PF of producer (i.e. number of PEs) does not match that of the consumer, two sets of buffers with data shuffling logic in between may be needed. For nodes that can complete their execution in linear time, e.g., matrix addition, this data shuffling eliminates any performance benefit gained by parallelizing its execution. 

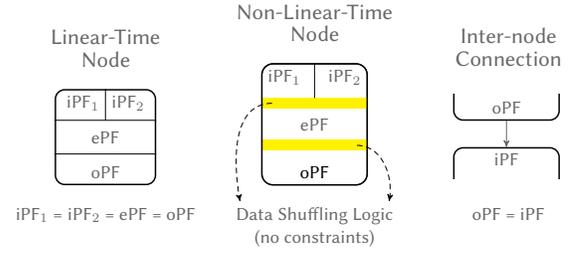
\begin{figure}[bt]
    \centering
    \scalebox{0.7}{\begin{tikzpicture}[font=\sffamily\fontfamily{LinuxBiolinumT-LF}\selectfont,>=stealth',text=black!60]

  \tikzset{linear/.style={align=center,draw,rounded corners=5pt,inner sep=5pt}};
  \tikzset{nonlinear/.style={align=center,draw,rounded corners=6pt,inner sep=5pt}};

  \node (l1) [linear,minimum width=19mm,minimum height=18mm,thick] {};
  \node at (l1.north west) [anchor=north west,xshift=1mm] {iPF$_1$};
  \node at (l1.north east) [anchor=north east,xshift=-1mm] {iPF$_2$};
  \node at (l1) {ePF};
  \node at (l1.south) [anchor=south] {oPF};
  
  \coordinate (nww) at ($(l1.north west)!0.65!(l1.west)$);
  
  \draw  (nww) -- ($(l1.north east)!0.65!(l1.east)$);
  \draw ($(l1.south west)!0.65!(l1.west)$) -- ($(l1.south east)!0.65!(l1.east)$);
  \draw (l1.north) -- (l1.north|-nww);


  
  \node (lconst) at ([yshift=-3mm]l1.south) [anchor=north] {
  iPF$_1$ = iPF$_2$ = ePF = oPF
  };
  
  \node at (l1.north) [anchor=south,yshift=3mm,align=center] {\large Linear-Time\\\large Node};

  
  \node (nl1) at (l1.south east) [nonlinear,anchor=south west,xshift=20mm,minimum height=23mm,minimum width=20mm] {};
  \node at (nl1.north west) [anchor=north west] {iPF$_1$};
  \node at (nl1.north east) [anchor=north east] {iPF$_2$};
  \node at (nl1) [] {ePF};
  \node at (nl1.south) [anchor=south,black] {oPF};
  \fill [yellow] ([yshift=5mm]nl1.west) rectangle ([yshift=3mm]nl1.east);
  \fill [yellow] ([yshift=-5mm]nl1.west) rectangle ([yshift=-3mm]nl1.east);
  \draw (nl1.north) -- ([yshift=5mm]nl1.center);
  
  \node at (nl1) [nonlinear,minimum height=23mm,minimum width=20mm,thick] {};


  
  \node (nlconst) at (lconst.north-|nl1.south) [anchor=north,align=center] {Data Shuffling Logic\\(no constraints)};
  
  \node at (nl1.north) [anchor=south,yshift=3mm,align=center] {\large Non-Linear-Time\\\large Node};
  
  \draw [->,densely dashed] ([yshift=4mm,xshift=2mm]nl1.west) to[out=180,in=100] ([xshift=2mm]nlconst.north west);
  
  \draw [->,densely dashed] ([yshift=-4mm,xshift=-2mm]nl1.east) to[out=0,in=80] ([xshift=-2mm]nlconst.north east);
  
  
  \node (connect) at (nl1.south east) [anchor=south west,xshift=15mm] {
  \begin{tikzpicture}
    \clip (0,0.5) rectangle (2.1,-1.1);
    
    \node (lc1) [linear,black,minimum height=15mm,minimum width=20mm,thick] {};
    \node (lc2) at (lc1.south) [linear,black,anchor=north,yshift=-5mm,minimum height=15mm,minimum width=20mm,thick] {};
    
    \node at (lc1.south) [anchor=south] {oPF};
    \node at (lc2.north) [anchor=north] {iPF};
    
    \draw [->] (lc1) -- (lc2);
  \end{tikzpicture}
  };
  
  \node at (lconst.north-|connect.south) [anchor=north] {oPF = iPF};
  
  \node at (connect.north) [anchor=south,yshift=3mm,align=center] {\large Inter-node\\\large Connection}; 
\end{tikzpicture}

        
        
        
        



}
    \caption{Linear-time and non-linear-time nodes}
    \label{fig:pf-constraints}
\end{figure}

To avoid this problem, we introduce additional constraints on PFs (shown in Figure~\ref{fig:pf-constraints}). We classify nodes as: 1)~\emph{linear-time nodes}, those that can complete execution in linear time (or less) in terms of the input size (e.g., matrix addition), and 2)~\emph{non-linear-time nodes}, those that take worse than linear time to complete execution (e.g., matrix multiplication). We associate each node with multiple PFs: one for each input, one for its execution unit, and one for its output. The figure shows a two-input linear-time node and a two-input non-linear-time node. For a linear-time node, we require the input, execution, and the output PFs to be the same (thereby avoiding the need for any data shuffling). For non-linear time nodes, we introduce a logic before and after the execution to appropriately shuffle the data. Finally, for two nodes with a producer-consumer relationship, the output PF of the producer must be equal to the input PF of the consumer. One natural implication of these constraints is that a sub-graph of connected linear-time nodes will all have the same PFs. We exploit this observation further to enable pipelined execution of linear-time nodes (Section~\ref{sec:pipelining}).

\subsection{Latency/Resource Estimation Models}
\label{sec:latency-resource-estimation}
The key optimization problem in \tool is to determine the best PF for each node in the DFG such that the overall latency is minimized, while fitting inside the resource budget of the FPGA. 
This entails exploration of a vast solution space of possible PF assignments. To reduce exploration time, we build regression models for each template that can predict the resource consumption and latency of a node. The overall resource consumption of a candidate solution is the sum of the resource consumption of all nodes and the execution latency is the sum of the latency of all the nodes in the \emph{critical} path (path with maximum latency). For our target models, we find that the buffering is done mainly using distributed RAM (LUTRAM) because of smaller matrix dimensions. As a result, there are typically more than enough memory resources (BRAM, Flip Flops) on our target FPGA boards, so we focus on predicting only the compute resources (LUT+LUTRAM, DSP).  

\label{sec:model-setup}

We build a separate estimation model for each type of matrix operation. 
Given the input dimensions of the operation and a target PF, our model estimates the resource consumption (LUTs and DSPs) and execution latency of the operation. 
In general, we expect LUT consumption to increase linearly with PF.
The DSP usage in a PE is known statically for an operation. We expect latency of a node, in general, to reduce linearly as $1/\textrm{PF}$. However, we find that for some operations, the parallel execution is followed by some linear reduction of the partial sums (e.g., \textsf{DotProduct}). Accordingly, we use the following models for estimating LUT, DSP, and latency consumption of a node.
In the equations below, LUT[1] and Latency[1] are the LUT consumption and latency of the corresponding operation with the specified input dimensions when the PF is set to 1. 
\begin{eqnarray*}
\textrm{Latency}[\textrm{PF}] & = & (\alpha_L + \beta_L \cdot \textrm{PF} + \gamma_L\cdot\frac{1}{\textrm{PF}})\cdot\textrm{Latency}[1] \\
\textrm{LUT}[\textrm{PF}] & = & (\alpha_{LUT} + \beta_{LUT}.\textrm{PF})\cdot \textrm{LUT}[1]\\
\textrm{DSP}[\textrm{PF}] & = & \alpha_{DSP} \cdot \textrm{PF}
\end{eqnarray*}


These equations require values for $\alpha_{LUT}$, $\beta_{LUT}$, $\alpha_{DSP}$, $\alpha_{L}$, $\beta_{L}$, and $\gamma_{L}$ for each matrix type.
Of these, $\alpha_{DSP}$ is set by the template developer to the number of DSPs used by the template's PE. We use a  training algorithm to find the remaining parameters.  To do this, we generate multiple sets of training data points. Within each set, we fix the input dimensions to an arbitrary value and vary the PF from 1 to a number beyond which the operation cannot be further parallelized by the underlying template. We synthesize and simulate Verilog implementations for each set to obtain the true LUT consumption and execution latency for each operation. We then identify the best parameters that minimize the mean squared error on the training data set. This is an one-time-effort for a family of FPGAs (Artix-7 in our case). These regression models are pre-trained during the tool development and are already included as a part of the \tool framework. 

\subsection{Data Flow Graph Generator}
\label{sec:dfg-generator}

\tool constructs the data flow graph (DFG) of the input program by analyzing the data dependencies in the program. Each node in the DFG is annotated with 1)~type of operation, 2)~input dimensions for the operation, and 3)~any static model parameters for the operation. \tool currently includes a DFG generator for \toolorig DSL~\cite{Seedot}. We also support a subset of Tensorflow~\cite{tensorflow} by converting the Tensorflow program to \toolorig and extracting the DFG.



\subsection{PF-1 Profiler}
\label{sec:pf1-profiler}
The \textit{PF-1 profiler} determines 
\textrm{LUT}[1] and \textrm{Latency}[1] described earlier. 
For each node in the program's DFG, it
synthesizes the Verilog implementation of the corresponding template with the node's operation dimensions to obtain the value of LUT[1] (once for each node). It then simulates the whole design with a random input to measure the PF=1 latency (one-time).
The \textit{PF-1 profiler} tags each node of the DFG with the measured values and passes the DFG to the next stage.

\subsection{Best-PF Estimator}
\label{sec:best-pf-estimator}

The goal of the Best-PF Estimator is to determine the best PF assignment for each node. The best PF of a node depends on the criticality of the node in the input program, which is influenced by the node's data dimensions. For example, across the 20 datasets we evaluate in Section~\ref{sec:results}, the PF for the SpMV node generated by the \tool optimizer ranges from 3 to 71. This further demonstrates the inefficiency of existing DNN accelerators that optimize for a single type of workload.

Since all the execution PFs are integers, the optimization problem can be expressed as an integer program. We explore two optimization strategies for the best-PF estimator.


\subsubsection{Black-box optimization}
\label{sec:lp-approximation}

In this approach we employ a generic solver to solve the optimization problem.
The critical path of the program can change as PF assignments change. Therefore, we express this optimization metric as a min-max problem: minimize the maximum latency across all paths in the program. 
Our framework adds constraints for each path in the DFG, stating that the sum of latency of all the nodes for each path should be less than a target latency. The integer program is then setup to minimize this target latency. Since solving an integer program is NP-hard, we relax it to be an optimization problem over real numbers and round the PFs to the nearest integer.

\subsubsection{Greedy Optimization}
\label{sec:greedy-optimization}

As finding a global optimum can be computationally intensive, we also explore a greedy strategy. Our greedy algorithm initializes the PFs of all the nodes to one. It then repeats the following steps while resources are still available.
\begin{enumerate}
    \item Identify the current critical path of the program.
    \item Identify the most critical node in that path. The most critical node is the one which yields the maximum \emph{benefit} when its PF is increased by one subject to the constraints established in Section~\ref{sec:inter-node-communication}. Note that when a node's PF is increased by one, the optimizer may also have to increase the PF of some connected nodes.
    \item If the optimizer cannot increase the PF of any node on the critical path, it exits immediately. As \tool executes the program in data flow order, there is no benefit to parallelizing a non-critical node even if there are resources available.
    \item If not, \tool increases the PF of the critical node by one and iterates.
\end{enumerate}
We support two \emph{benefit} metrics: reduction in latency and reduction in latency per additional LUT consumed.

\subsection{Scheduler Generator}
\label{sec:scheduler}

\tool enables concurrent execution of nodes wherever possible. 
Each template is associated with a \emph{start} and \emph{done} signal. A node can start execution as soon as its \emph{start} signal is asserted. When a node completes execution, it asserts its \emph{done} signal. \tool generates the controller that encodes this scheduling logic for the program's DFG.



\subsection{Pipelining Linear-Time Nodes}
\label{sec:pipelining}

When consecutive operations have the same PF, it is possible to 
view these two nodes as a super node and pipeline their execution (rather than waiting for the first to complete before starting the second).
This optimization eliminates the need for memory buffers between pipelined nodes. To perform this optimization, \tool identifies clusters of linear-time nodes that are connected to each other and pipelines them. The pipeline begins execution only when \emph{all} the nodes supplying input to the pipeline have completed execution.

\section{Evaluation Methodology}
\label{sec:eval_method}

    
\subsection{Hardware Setup and Benchmarks}
We target the small form-factor Xilinx Arty-board, which has 20800 logic units (LUTs), 90 DSP (multiplier) slices and 225~KB of on-chip memory operating at 10MHz. For comparison with microcontrollers, we use the results presented by~\toolorig~\cite{Seedot}. \toolorig targets the Arduino Uno Board (8-bit ATmega328P) with 2KB of SRAM and 32KB of read-only flash operating at 16MHz.


We evaluate different tools using two state-of-the-art machine learning algorithms: \bonsai~\cite{bonsai}, a decision-tree based classifier, and \protonn~\cite{protonn}, a k-nearest-neighbour based classifier, from Microsoft's EdgeML library~\cite{edgeml}. These algorithms are crafted specifically to run on milliwatt-scale IoT devices and have state-of-the-art accuracies on various tasks. 
We use ten standard ML datasets (both binary and multiclass): cifar~\cite{cifar}, character recognition (cr)~\cite{cr}, usps~\cite{usps}, mnist~\cite{mnist}, letter~\cite{letter} and ward~\cite{ward}. In total, we evaluate 20 different DFGs as our benchmarks. 

\subsection{Comparison Points}
\label{sec:comparison-points}

\begin{figure*}[t]
\centering
\scalebox{0.75}{\input{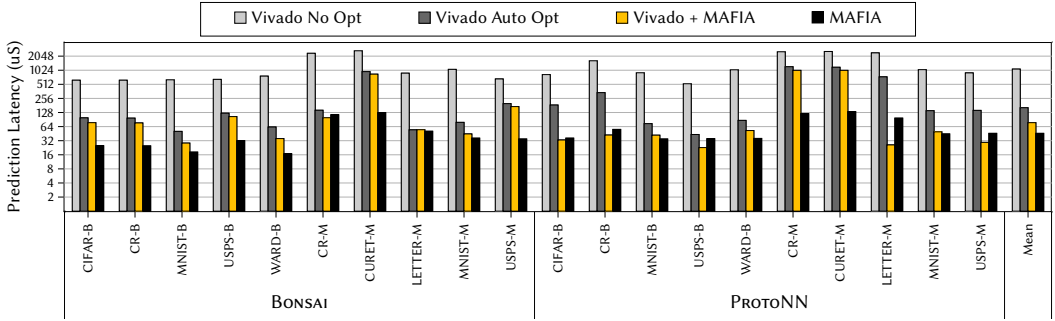}}
\caption{Prediction latency using different compilers (lower is better). Y-axis is in log-scale.}
\label{plot:latency}
\end{figure*}

We compare \tool with four mechanisms that can currently be used for generating Verilog programs from high-level specifications for our target models.
We leverage the quantization scheme from a prior work, \toolorig~\cite{Seedot}. The fixed-point programs generated from \toolorig are used in our evaluations.\\
\noindent \textbf{Bambu HLS and Vivado HLS}
We run Bambu with the \texttt{BAMBU-PERFORMANCE-MP} flag (\texttt{-O3} optimizations) and Vivado HLS with the default options without any additional compiler hints (denoted by \vivado. in Figure~\ref{plot:latency}).

\noindent \textbf{\vivadoauto.}
In this variant, we compare our work against a recent work, \toolorig~\cite{Seedot}. \toolorig is a language and a compiler to generate efficient implementation of ML inference algorithms targeting low-end IoT devices. The FPGA backend of \toolorig 1)~uses a hand-optimized Verilog implementation for Sparse Matrix Vector Multiplication (the most time consuming kernel in the microcontroller-based implementations of~\cite{protonn,bonsai}) with a fixed parallelism factor of 10 and 2)~accelerates the rest of the program by automatically annotating the input C program to Vivado HLS with loop unrolling and pipelining hints.


\noindent \textbf{\vivadoplus.}
To understand the best we could do using a commercial HLS tool, we use the parallelism factors obtained from the \tool optimizer to generate appropriate hints to the Vivado HLS compiler. We build this variant on top of \vivadoauto. First, we set the parallelism factor for the hand-optimized SpMV implementation to the PF value of the SpMV node generated by the \tool optimizer. Second, we iteratively incorporate the other PFs from the \tool optimizer by appropriately unrolling the outermost loop in the C-code of the corresponding operation.
%
%
We then manually improve the design performance by further unrolling the loops of the program, until the solution runs out of resource budget. 

\noindent \textbf{\mafia.}
In our evaluations, we extract the DFGs from the \toolorig~\cite{Seedot} framework and use them as input for \tool. All our main results are presented with the best performing configuration in which the code is generated by our greedy optimizer (see Section~\ref{sec:greedy-vs-convex} for more details) with latency reduction per additional LUT as the benefit metric.


\section{Results}
\label{sec:results}

\begin{figure}[b]
    \centering
    \scalebox{0.9}{\input{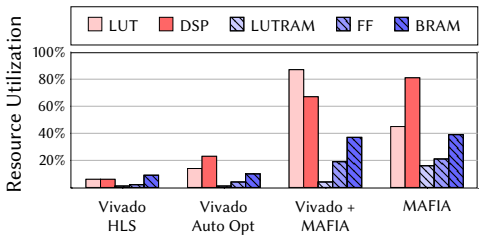}}
    \caption{FPGA resource utilization}
    \label{plot:resource-results}
\end{figure}

We first compare \tool-generated programs with other prior approaches on both prediction latency and resource utilization (Section~\ref{sec:performance-results}). Our results show that \tool generates programs that reduce prediction latency by 2.5$\times$ on average compared to the best previous approach. Figure~\ref{plot:resource-results} illustrates the efficiency of \tool in allocating resources based on criticality of nodes. This is underlined in our evaluation in Section~\ref{sec:regression-model-accuracy}, where we observe that our approach is more accurate than existing tools.  Finally, in Section~\ref{sec:greedy-vs-convex}, our evaluation of two optimization strategies, greedy and black-box optimization, reveal that the greedy approach is significantly faster and generates programs of similar quality (if not better) compared to the black-box approach.


\subsection{Comparison of Different Mechanisms}
\label{sec:performance-results}

Our results show that all FPGA tools perform comfortably better than programs running on microcontrollers. \vivado. performs 14$\times$ better than microcontroller implementations.  However, as expected \tool outperforms both Bambu HLS and \vivado significantly. 
Therefore, in the rest of the paper, we present results on the following mechanisms: 1)~\vivadoauto, 2)~\vivadoplus, and 3)~\mafia.

Figure~\ref{plot:latency} plots the prediction latency of these mechanisms for all twenty benchmarks. The y-axis is in log-scale. Table~\ref{table-baseline-latency} lists the number of features for each dataset and their baseline (microcontroller) latencies. Figure~\ref{plot:resource-results} shows the average utilization of LUT, DSP, LUT RAM, Flip Flops, and Block RAMs.  Since we find memory resources not to be a bottleneck, we focus our attention on LUT and DSP.

\subsubsection{\vivadoauto.}

\vivadoauto improves prediction latency by 7$\times$ on average and quadruples the utilization of compute resources compared to \vivado. by using a hand-optimized implementation for SpMV and adding automatic loop unrolling hints.
However, this prior approach is still lacking in two aspects. First, it uses a fixed parallelism factor of 10 for SpMV on all data sets. But, we observe that the criticality of this operation varies significantly with different data sets. As detailed in Section~\ref{sec:best-pf-estimator}, the PF for SpMV generated by the \tool optimizer ranges from 3 to 71. Second, the mechanism used by prior work to estimate resource utilization had high error rate and results in subpar loop unrolling hints.

\subsubsection{\vivadoplus.}  

\vivadoplus involves two steps. First, we incorporate the PFs generated by \tool's optimizer in the code generated by \vivadoauto by appropriately setting the PF of the SpMV node and the loop unrolling factors for the other nodes. Although, this improved the performance by ~27\% on average, we observed that this approach still resulted in suboptimal performance. This is because, \emph{Best-PF Estimator} of \tool assumes that the program will be executed in data flow order. However, Vivado HLS does not execute independent nodes in parallel. Therefore, even parallelizing the non-critical nodes in the DFG can enable better performance. We manually unrolled loops for non-critical nodes as well till we hit the resource budget. With this improvement, \vivadoplus outperforms \vivadoauto by up to 3.5$\times$ on average across all datasets.

\begin{table}[b]
\centering
\scalebox{0.8}{
\small
{
 \begin{tabular}{ c c  c  c}
 \toprule

\multirow{2}{*}{\textbf{DATASET}} & \multirow{2}{*}{\textbf{\parbox[t]{1.5cm}{\centering Num\\Features}}} & \bonsai & \protonn\\

 & & \parbox[t]{1.5cm}{\centering Baseline\\Latency(us)} & \parbox[t]{1.5cm}{\centering Baseline\\Latency(us)}\\
\toprule
cifar-b & 400 & 6121 & 14112\\ 

cr-b & 400  & 6263 & 28446\\

mnist-b & 784 & 11568 & 15983 \\ 

usps-b & 256 & 4099 & 9206\\

ward-b & 1000 & 14733 & 23241\\

cr-m & 400 & 29030 & 34667\\

curet-m & 610 & 39731 & 37769\\ 

letter-m & 16 & 11161 & 35377\\

mnist-m & 784 & 16026 & 18491\\

usps-m & 256 & 9140 & 14017\\
 
\bottomrule
\end{tabular}
}
}
\caption{\centering Baseline (microcontroller) latency and number of features for the evaluated datasets}
\label{table-baseline-latency}
\end{table}

\subsubsection{\mafia.}

Our analysis of the performance and resource consumption of the different mechanisms show that \mafia improves upon the state-of-the-art on two fronts. First, \tool is able to better estimate required parallelism factors for each node. Second, \tool executes the program in data flow order, rather than sequentially. \mafia outperforms \vivadoauto by 4.2$\times$ and even \vivadoplus (with manual hints) by 2.5$\times$ despite consuming only half the LUT resources compared to \vivadoplus. This highlights the fact that the performance improvement in C-based HLS tools is only due to intra-node parallelism. With static knowledge of the program's data flow graph, \tool is able to schedule operations in data flow order, and thereby execute independent nodes in parallel. Unfortunately, there are no simple hints that can be provided to Vivado HLS to express this opportunity. Also, the hand-optimized implementation of each matrix operation template allows \tool to more efficiently perform the underlying arithmetic operations.
Therefore, \tool exploits both inter-node and intra-node parallelism and can offer higher benefits on more resource constrained FPGAs.

\subsection{\tool Resource/Latency Estimation Models}
\label{sec:regression-model-accuracy}
The accuracy of \tool's estimation models ensures that the optimizer does not overestimate resources and yields suboptimal results by setting lower PFs. The final error of \tool estimation models is 36\% for LUT, 17\% for DSPs, and 99\% for overall latency. The high error in latency estimate is in part due to the fact that our estimation algorithm does not capture the pipelining optimization (Section~\ref{sec:pipelining}). Pipelining results  in significant reduction in latency while consuming similar amount of compute resources. However, the latency model correctly captures the relative latencies of all nodes, which is sufficient to guide the \tool-optimizer. In any case, to put these numbers in context, the estimation error of the Vivado HLS compiler for programs generated by \vivadoplus is 73\% for LUT, 673\% for DSP, and fails to provide latency estimates.


\subsection{Greedy vs Black-box optimization}
\label{sec:greedy-vs-convex}

So far, we have presented results using our greedy optimization strategy. As described in Section~\ref{sec:lp-approximation}, we also explore the use of a generic solver. We observe that the prediction latency of programs generated by the greedy approach are on average 10\% \emph{lower} than those obtained from the black-box approach (for \bonsai across all data sets). The greedy optimizer is also \emph{significantly} faster than the black-box approach (22$\times$ on average). The better performance of the greedy approach can be attributed to our rounding mechanism in the black-box optimization. To ensure that we fit within the resource budget of the FPGA, we round down all the PF numbers obtained from the black-box solver. Optimal rounding is itself an NP-hard problem.

\textbf{Power Consumption:}~The average power consumption across all implementations (20 datasets) is 76.15 mW.

\section{Related Work}
\label{sec:related}

Development of small, accurate ML models for IoT devices is a budding research area~\cite{bonsai, protonn, emirnn, fastgrnn} with several novel and interesting applications including face-detection, assistance for visual impaired, remote farming and radar classification. We believe \tool can allow these applications to be run on similar form-factor FPGA instead of microcontrollers, thereby providing significant boost in the performance of the underlying models. Many \emph{ML-HLS tools}~\cite{leflow,TF2FPGA,hls4ml,fp-dnn,dnnweaver,autodnn,brainwave,fastwave} exist to compile ML inference to FPGAs from popular frameworks like Tensorflow and Pytorch. Unfortunately, all of them target DNN workloads and focus their attention on high-end FPGAs~\cite{cnn1, rnn1, accOp1, accOp2}. 
Other works on automatically extracting parallelism in general-purpose HLS tools include~\cite{affine1, affine2, affine3}. The closest work to \tool is \toolorig~\cite{Seedot}, a framework that compiles efficient code for microcontrollers and low-end FPGAs from a high-level specification.

\section{Conclusion}
\label{sec:conclusion}

ML inference models targeting milli-watt powered IoT devices depart from the DNN architecture and use sophisticated primitives with fewer parameters~\cite{protonn,bonsai}.
However, most current HLS for ML research focuses on neural networks and target high-end FPGAs, resulting in tools that lack the requisite expressiveness. Thus, developers in the IoT space have to rely on general purpose HLS tools that have sufficient expressiveness but suffer from poor resource utilization. \tool achieves both high expressiveness and efficiency; it is a fully automatic toolchain that compiles device-agnostic ML algorithms to efficient Verilog code. We have evaluated MAFIA on state-of-the-art ML algorithms in the IoT space and it outperforms prior work by $2.5\times$ on average.  

\section{Acknowledgements}
This work was funded by NSERC through the COHESA strategic network.



\printbibliography

@inproceedings{Seedot,
 author = {Gopinath, Sridhar and Ghanathe, Nikhil and Seshadri, Vivek and Sharma, Rahul},
 title = {{Compiling KB-sized Machine Learning Models to Tiny IoT Devices}},
 booktitle = {Proceedings of the 40th ACM SIGPLAN Conference on Programming Language Design and Implementation},
 series = {PLDI 2019},
 year = {2019},
 isbn = {978-1-4503-6712-7},
 location = {Phoenix, AZ, USA},
 pages = {79--95},
 numpages = {17},
 %url = {http://doi.acm.org/10.1145/3314221.3314597},
 %doi = {10.1145/3314221.3314597},
 %acmid = {3314597},
 %publisher = {ACM},
 %address = {New York, NY, USA},
 %keywords = {Compiler, FPGA, Fixed-point, IoT device, Machine Learning, Microcontroller, Programming Language},
}

@book{mitchell,
  author    = {Tom M. Mitchell},
  title     = {Machine learning, International Edition},
  series    = {McGraw-Hill Series in Computer Science},
  publisher = {McGraw-Hill},
  year      = {1997},
  url       = {https://www.worldcat.org/oclc/61321007},
  isbn      = {978-0-07-042807-2},
}

@inproceedings{protonn,
	title={{ProtoNN: Compressed and accurate kNN for resource-scarce devices}},
	author={Gupta, Chirag and Suggala, Arun Sai and Goyal, Ankit and Simhadri, Harsha Vardhan and Paranjape, Bhargavi and Kumar, Ashish and Goyal, Saurabh and Udupa, Raghavendra and Varma, Manik and Jain, Prateek},
	booktitle={International Conference on Machine Learning},
	pages={1331--1340},
	year={2017}
}

@inproceedings{bonsai,
	title={{Resource-efficient Machine Learning in 2 KB RAM for the Internet of Things}},
	author={Kumar, Ashish and Goyal, Saurabh and Varma, Manik},
	booktitle={International Conference on Machine Learning},
	pages={1935--1944},
	year={2017}
}

@techreport{cifar,
	title={{Learning multiple layers of features from tiny images}},
	author={Krizhevsky, Alex},
	year={2009},
	institution={Citeseer}
}

@inproceedings{cr,
	author    = {Te{\'{o}}filo Em{\'{\i}}dio de Campos and Bodla Rakesh Babu and Manik Varma},
	title     = {{Character Recognition in Natural Images}},
	booktitle = {{VISAPP} 2009 - Proceedings of the Fourth International Conference on Computer Vision Theory and Applications, Lisboa, Portugal, 2009, Vol. 2},
	pages     = {273--280},
	year      = {2009},
}

@article{letter,
	title={{A comparison of methods for multiclass support vector machines}},
	author={Hsu, Chih-Wei and Lin, Chih-Jen},
	journal={IEEE transactions on Neural Networks},
	volume={13},
	number={2},
	pages={415--425},
	year={2002},
	publisher={IEEE}
}

@article{mnist,
	title={{Gradient-based learning applied to document recognition}},
	author={LeCun, Yann and Bottou, L{\'e}on and Bengio, Yoshua and Haffner, Patrick},
	journal={Proceedings of the IEEE},
	volume={86},
	number={11},
	pages={2278--2324},
	year={1998},
	publisher={IEEE}
}

@article{usps,
	title={{A database for handwritten text recognition research}},
	author={Hull, Jonathan J.},
	journal={IEEE Transactions on pattern analysis and machine intelligence},
	volume={16},
	number={5},
	pages={550--554},
	year={1994},
	publisher={IEEE}
}

@inproceedings{ward,
	title={{Group-sensitive multiple kernel learning for object categorization}},
	author={Yang, Jingjing and Li, Yuanning and Tian, Yonghong and Duan, Lingyu and Gao, Wen},
	booktitle={Computer Vision, 2009 IEEE 12th International Conference on},
	pages={436--443},
	year={2009},
	organization={IEEE}
}

@inproceedings{gesturepod,
author = {Patil, Shishir G. and Dennis, Don Kurian and Pabbaraju, Chirag and Shaheer, Nadeem and Simhadri, Harsha Vardhan and Seshadri, Vivek and Varma, Manik and Jain, Prateek},
title = {GesturePod: Enabling On-Device Gesture-Based Interaction for White Cane Users},
year = {2019},
isbn = {9781450368162},
publisher = {Association for Computing Machinery},
address = {New York, NY, USA},
url = {https://doi.org/10.1145/3332165.3347881},
doi = {10.1145/3332165.3347881},
booktitle = {Proceedings of the 32nd Annual ACM Symposium on User Interface Software and Technology},
pages = {403–415},
numpages = {13},
location = {New Orleans, LA, USA},
series = {UIST '19}
}

@inproceedings{dnnweaver,
	author = {Sharma, Hardik and Park, Jongse and Mahajan, Divya and Amaro, Emmanuel and Kim, Joon Kyung and Shao, Chenkai and Mishra, Asit and Esmaeilzadeh, Hadi},
	title = {{From High-level Deep Neural Models to FPGAs}},
	booktitle = {The 49th Annual IEEE/ACM International Symposium on Microarchitecture},
	series = {MICRO-49},
	year = {2016},
	location = {Taipei, Taiwan},
	pages = {17:1--17:12},
	articleno = {17},
	numpages = {12},
%	acmid = {3195659},
%	publisher = {IEEE Press},
%	address = {Piscataway, NJ, USA},
}

@inproceedings{fp-dnn, 
	author={Y. Guan and H. Liang and N. Xu and W. Wang and S. Shi and X. Chen and G. Sun and W. Zhang and J. Cong}, 
	booktitle={2017 IEEE 25th Annual International Symposium on Field-Programmable Custom Computing Machines (FCCM)}, 
	title={{FP-DNN: An Automated Framework for Mapping Deep Neural Networks onto FPGAs with RTL-HLS Hybrid Templates}}, 
	year={2017}, 
	volume={}, 
	number={}, 
	pages={152-159}, 
	doi={10.1109/FCCM.2017.25}, 
	ISSN={}, 
	month={April},
}

@InProceedings{farmbeats,
author = {Vasisht, Deepak and Kapetanovic, Zerina and Won, JongHo and Jin, Xinxin and Chandra, Ranveer and Sinha, Sudipta and Kapoor, Ashish},
title = {{FarmBeats: An IoT Platform for Data-Driven Agriculture}},
booktitle = {Networked Systems Design and Implementation (NSDI)},
year = {2017},
month = {March},
publisher = {USENIX},
url = {https://www.microsoft.com/en-us/research/publication/farmbeats-iot-platform-data-driven-agriculture/},
edition = {Networked Systems Design and Implementation (NSDI)},
}

@article{cloud1,
	title={{A cloud-based car parking middleware for IoT-based smart cities: Design and implementation}},
	author={Ji, Zhanlin and Ganchev, Ivan and O'Droma, M{\'a}irt{\'\i}n and Zhao, Li and Zhang, Xueji},
	journal={Sensors},
	volume={14},
	number={12},
	pages={22372--22393},
	year={2014},
	publisher={Multidisciplinary Digital Publishing Institute}
}

@article{cloud2,
	title={{Internet of Things (IoT): A vision, architectural elements, and future directions}},
	author={Gubbi, Jayavardhana and Buyya, Rajkumar and Marusic, Slaven and Palaniswami, Marimuthu},
	journal={Future generation computer systems},
	volume={29},
	number={7},
	pages={1645--1660},
	year={2013},
	publisher={Elsevier}
}

@inproceedings{cloud3,
	title={{Health monitoring and management using Internet-of-Things (IoT) sensing with cloud-based processing: Opportunities and challenges}},
	author={Hassanalieragh, Moeen and Page, Alex and Soyata, Tolga and Sharma, Gaurav and Aktas, Mehmet and Mateos, Gonzalo and Kantarci, Burak and Andreescu, Silvana},
	booktitle={2015 IEEE international conference on services computing (SCC)},
	pages={285--292},
	year={2015},
	organization={IEEE}
}

@inproceedings{brainwave, 
	author={J. Fowers and {et. al.}
	},
	booktitle={2018 ACM/IEEE 45th Annual International Symposium on Computer Architecture (ISCA)}, 
	title={{A Configurable Cloud-Scale DNN Processor for Real-Time AI}}, 
	year={2018}, 
	volume={}, 
	number={}, 
	pages={1-14}, 
	doi={10.1109/ISCA.2018.00012}, 
	ISSN={2575-713X}, 
	month={June},
}

@INPROCEEDINGS{rnn1,  author={Y. {Sun} and B. {Liu} and X. {Xu}},  booktitle={2019 International Conference on Field-Programmable Technology (ICFPT)},   title={{An OpenCL-Based Hybrid CNN-RNN Inference Accelerator On FPGA}},   year={2019},  volume={},  number={},  pages={283-286},}

@software{edgeml,
   author = {{Dennis, Don Kurian and Gopinath, Sridhar and Gupta, Chirag and Kumar, Ashish and Kusupati, Aditya and Patil, Shishir G and Simhadri, Harsha Vardhan}},
   title = {{EdgeML: Machine Learning for resource-constrained edge devices}},
   url = {https://github.com/Microsoft/EdgeML},
   version = {0.1},
}

@inproceedings{fastgrnn,
  author = {Aditya Kusupati and Manish Singh and Kush Bhatia and Ashish Kumar and Prateek Jain and Manik Varma},
  title = {{FastGRNN: A Fast, Accurate, Stable and Tiny Kilobyte Sized Gated Recurrent Neural Network}},
  booktitle = {Proceedings of the Thirty-first Annual Conference on Neural Information Processing Systems (NeurIPS)},
  year = {2018},
  pages = {9031--9042},
  note = {slides/fastgrnn.pdf},
  url = {all_papers/KusupatiSBKJV18.pdf}
}

@inproceedings{emirnn,
  author = {Don Dennis and Chirag Pabbaraju and Harsha Vardhan Simhadri and Prateek Jain},
  title = {{Multiple Instance Learning for Efficient Sequential Data Classification on Resource-constrained Devices}},
  booktitle = {Proceedings of the Thirty-first Annual Conference on Neural Information Processing Systems (NeurIPS)},
  year = {2018},
  pages = {10976--10987},
  note = {slides/DennisPSJ18.pdf},
  url = {all_papers/DennisPSJ18.pdf}
}

@INPROCEEDINGS{autodnn,  author={X. {Zhang} and J. {Wang} and C. {Zhu} and Y. {Lin} and J. {Xiong} and W. {Hwu} and D. {Chen}},  booktitle={2018 IEEE/ACM International Conference on Computer-Aided Design (ICCAD)},   title={{DNNBuilder: an Automated Tool for Building High-Performance DNN Hardware Accelerators for FPGAs}},   year={2018},  volume={},  number={},  pages={1-8},}

@ARTICLE{cnn1,  author={X. {Lian} and Z. {Liu} and Z. {Song} and J. {Dai} and W. {Zhou} and X. {Ji}},  journal={IEEE Transactions on Very Large Scale Integration (VLSI) Systems},   title={{High-Performance FPGA-Based CNN Accelerator With Block-Floating-Point Arithmetic}},   year={2019},  volume={27},  number={8},  pages={1874-1885},}

@INPROCEEDINGS{leflow,  
    author={D. {Holanda Noronha} and K. {Gibson} and B. {Salehpour} and S. J. E. {Wilton}},  
    booktitle={2018 International Conference on Field-Programmable Technology (FPT)},   
    title={{LeFlow: Automatic Compilation of TensorFlow Machine Learning Applications to FPGAs}},   
    year={2018},  
    volume={},  
    number={},  
    pages={393-396}
}

@INPROCEEDINGS{TF2FPGA,  author={S. {Mouselinos} and V. {Leon} and S. {Xydis} and D. {Soudris} and K. {Pekmestzi}},  booktitle={2019 8th International Conference on Modern Circuits and Systems Technologies (MOCAST)},   title={{TF2FPGA: A Framework for Projecting and Accelerating Tensorflow CNNs on FPGA Platforms}},   year={2019},  volume={},  number={},  pages={1-4},}

@article{fastwave,
   title={{FastWave: Accelerating Autoregressive Convolutional Neural Networks on FPGA}},
   ISBN={9781728123509},
   url={http://dx.doi.org/10.1109/ICCAD45719.2019.8942122},
   DOI={10.1109/iccad45719.2019.8942122},
   journal={IEEE/ACM International Conference on Computer-Aided Design (ICCAD)},
   publisher={IEEE},
   author={Hussain, Shehzeen and Javaheripi, Mojan and Neekhara, Paarth and Kastner, Ryan and Koushanfar, Farinaz},
   year={2019},
   month={}
}

@ARTICLE{hls4ml,
       author = {{Duarte}, J. and {Han}, S. and {Harris}, P. and {Jindariani}, S. and
         {Kreinar}, E. and {Kreis}, B. and {Ngadiuba}, J. and {Pierini}, M. and
         {Rivera}, R. and {Tran}, N. and {Wu}, Z.},
        title = {{Fast inference of deep neural networks in FPGAs for particle physics}},
      journal = {Journal of Instrumentation},
         year = 2018,
        month = jul,
       volume = {13},
       number = {7},
        pages = {P07027},
          doi = {10.1088/1748-0221/13/07/P07027},
archivePrefix = {arXiv},
       eprint = {1804.06913},
 primaryClass = {physics.ins-det},
       adsurl = {https://ui.adsabs.harvard.edu/abs/2018JInst..13P7027D}
}

@article{decisiontree,
    author = {Quinlan, J. R.},
    title = {{Induction of Decision Trees}},
    year = {1986},
    issue_date = {March 1986},
    publisher = {Kluwer Academic Publishers},
    address = {USA},
    volume = {1},
    number = {1},
    issn = {0885-6125},
    url = {https://doi.org/10.1023/A:1022643204877},
    doi = {10.1023/A:1022643204877},
    journal = {Mach. Learn.},
    month = mar,
    pages = {81–106},
    numpages = {26}
}

@ARTICLE{kNN,
  author={T. {Cover} and P. {Hart}},
  journal={IEEE Transactions on Information Theory}, 
  title={{Nearest neighbor pattern classification}}, 
  year={1967},
  volume={13},
  number={1},
  pages={21-27},
}

@article{tensorflow,
  author    = {Mart{\'{\i}}n Abadi and
               Ashish Agarwal and
               Paul Barham and
               Eugene Brevdo and
               Zhifeng Chen and
               Craig Citro and
               Gregory S. Corrado and
               Andy Davis and
               Jeffrey Dean and
               Matthieu Devin and
               Sanjay Ghemawat and
               Ian J. Goodfellow and
               Andrew Harp and
               Geoffrey Irving and
               Michael Isard and
               Yangqing Jia and
               Rafal J{\'{o}}zefowicz and
               Lukasz Kaiser and
               Manjunath Kudlur and
               Josh Levenberg and
               Dan Man{\'{e}} and
               Rajat Monga and
               Sherry Moore and
               Derek Gordon Murray and
               Chris Olah and
               Mike Schuster and
               Jonathon Shlens and
               Benoit Steiner and
               Ilya Sutskever and
               Kunal Talwar and
               Paul A. Tucker and
               Vincent Vanhoucke and
               Vijay Vasudevan and
               Fernanda B. Vi{\'{e}}gas and
               Oriol Vinyals and
               Pete Warden and
               Martin Wattenberg and
               Martin Wicke and
               Yuan Yu and
               Xiaoqiang Zheng},
  title     = {TensorFlow: Large-Scale Machine Learning on Heterogeneous Distributed
               Systems},
  journal   = {CoRR},
  volume    = {abs/1603.04467},
  year      = {2016},
  url       = {http://arxiv.org/abs/1603.04467},
  archivePrefix = {},
  eprint    = {1603.04467},
  bibsource = {dblp computer science bibliography, https://dblp.org}
}

@inproceedings{accOp1,
author = {Hardieck, Martin and Kumm, Martin and M\"{o}ller, Konrad and Zipf, Peter},
title = {{Reconfigurable Convolutional Kernels for Neural Networks on FPGAs}},
year = {2019},
isbn = {9781450361378},
publisher = {Association for Computing Machinery},
address = {New York, NY, USA},
url = {https://doi.org/10.1145/3289602.3293905},
doi = {10.1145/3289602.3293905},
booktitle = {Proceedings of the 2019 ACM/SIGDA International Symposium on Field-Programmable Gate Arrays},
pages = {43–52},
numpages = {10},
location = {Seaside, CA, USA},
series = {FPGA '19}
}

@inproceedings{accOp2,
author = {Moss, Duncan J.M and Krishnan, Srivatsan and Nurvitadhi, Eriko and Ratuszniak, Piotr and Johnson, Chris and Sim, Jaewoong and Mishra, Asit and Marr, Debbie and Subhaschandra, Suchit and Leong, Philip H.W.},
title = {{A Customizable Matrix Multiplication Framework for the Intel HARPv2 Xeon+FPGA Platform: A Deep Learning Case Study}},
year = {2018},
%isbn = {9781450356145},
%publisher = {Association for Computing Machinery},
%address = {New York, NY, USA},
%url = {https://doi.org/10.1145/3174243.3174258},
%doi = {10.1145/3174243.3174258},
booktitle = {Proceedings of the 2018 ACM/SIGDA International Symposium on Field-Programmable Gate Arrays},
pages = {107–116},
numpages = {10},
keywords = {deep learning, FPGA, heterogeneous architectures, neural networks, reduced precision},
location = {Monterey, CALIFORNIA, USA},
series = {FPGA '18}
}

@inproceedings{affine1,
author = {Zuo, Wei and Liang, Yun and Li, Peng and Rupnow, Kyle and Chen, Deming and Cong, Jason},
title = {Improving High Level Synthesis Optimization Opportunity through Polyhedral Transformations},
year = {2013},
%isbn = {9781450318877},
%publisher = {Association for Computing Machinery},
%address = {New York, NY, USA},
%url = {https://doi.org/10.1145/2435264.2435271},
%doi = {10.1145/2435264.2435271},
abstract = {High level synthesis (HLS) is an important enabling technology for the adoption of hardware accelerator technologies. It promises the performance and energy efficiency of hardware designs with a lower barrier to entry in design expertise, and shorter design time. State-of-the-art high level synthesis now includes a wide variety of powerful optimizations that implement efficient hardware. These optimizations can implement some of the most important features generally performed in manual designs including parallel hardware units, pipelining of execution both within a hardware unit and between units, and fine-grained data communication. We may generally classify the optimizations as those that optimize hardware implementation within a code block (intra-block) and those that optimize communication and pipelining between code blocks (inter-block). However, both optimizations are in practice difficult to apply. Real-world applications contain data-dependent blocks of code and communicate through complex data access patterns. Existing high level synthesis tools cannot apply these powerful optimizations unless the code is inherently compatible, severely limiting the optimization opportunity. In this paper we present an integrated framework to model and enable both intra- and inter-block optimizations. This integrated technique substantially improves the opportunity to use the powerful HLS optimizations that implement parallelism, pipelining, and fine-grained communication. Our polyhedral model-based technique systematically defines a set of data access patterns, identifies effective data access patterns, and performs the loop transformations to enable the intra- and inter-block optimizations. Our framework automatically explores transformation options, performs code transformations, and inserts the appropriate HLS directives to implement the HLS optimizations. Furthermore, our framework can automatically generate the optimized communication blocks for fine-grained communication between hardware blocks. Experimental evaluation demonstrates that we can achieve an average of 6.04X speedup over the high level synthesis solution without our transformations to enable intra- and inter-block optimizations.},
booktitle = {Proceedings of the ACM/SIGDA International Symposium on Field Programmable Gate Arrays},
pages = {9–18},
numpages = {10},
keywords = {high level synthesis, polyhedral, FPGA},
location = {Monterey, California, USA},
series = {FPGA '13}
}

@INPROCEEDINGS{affine2,  author={G. {Natale} and G. {Stramondo} and P. {Bressana} and R. {Cattaneo} and D. {Sciuto} and M. D. {Santambrogio}},  booktitle={2016 IEEE/ACM International Conference on Computer-Aided Design (ICCAD)},   title={A polyhedral model-based framework for dataflow implementation on FPGA devices of Iterative Stencil Loops},   year={2016},  volume={},  number={},  pages={1-8},}

@ARTICLE{affine3,  author={J. {Liu} and J. {Wickerson} and S. {Bayliss} and G. A. {Constantinides}},  journal={IEEE Transactions on Computer-Aided Design of Integrated Circuits and Systems},   title={Polyhedral-Based Dynamic Loop Pipelining for High-Level Synthesis},   year={2018},  volume={37},  number={9},  pages={1802-1815},}

@INPROCEEDINGS{deepburning,  author={Y. {Wang} and J. {Xu} and Y. {Han} and H. {Li} and X. {Li}},  booktitle={53nd ACM/EDAC/IEEE Design Automation Conference},   title={DeepBurning: Automatic generation of FPGA-based learning accelerators for Neural Network family},   year={2016},  volume={},  number={},  pages={1-6},  doi={10.1145/2897937.2898002}}

@ARTICLE{reuse,  author={N. K. {Jha} and S. {Mittal}},  journal={IEEE Transactions on Computers},   title={Modeling Data Reuse in Deep Neural Networks by Taking Data-Types into Cognizance},   year={2020},  volume={},  number={},  pages={1-1},  doi={10.1109/TC.2020.3015531}}

@INPROCEEDINGS{ota_1,  author={T. {Gomes} and F. {Salgado} and S. {Pinto} and J. {Cabral} and A. {Tavares}},  booktitle={2016 IEEE 21st International Conference on Emerging Technologies and Factory Automation (ETFA)},   title={Towards an FPGA-based network layer filter for the Internet of Things edge devices},   year={2016},  volume={},  number={},  pages={1-4},  doi={10.1109/ETFA.2016.7733684}}

@inproceedings{ota_2,
author = {Jung, Jinhwan and Ryoo, Jihoon and Yi, Yung and Kim, Song Min},
title = {Gateway over the Air: Towards Pervasive Internet Connectivity for Commodity IoT},
year = {2020},
isbn = {9781450379540},
publisher = {Association for Computing Machinery},
address = {New York, NY, USA},
url = {https://doi.org/10.1145/3386901.3388949},
doi = {10.1145/3386901.3388949},
booktitle = {Proceedings of the 18th International Conference on Mobile Systems, Applications, and Services},
pages = {54–66},
numpages = {13},
location = {Toronto, Ontario, Canada},
series = {MobiSys '20}
}

@INPROCEEDINGS{reconfig-on-edge-1,  author={G. {Korol} and M. G. {Jordan} and M. {Brandalero} and M. {Hübner} and M. {Beck Rutzig} and A. C. {Schneider Beck}},  booktitle={2020 30th International Conference on Field-Programmable Logic and Applications (FPL)},   title={MCEA: A Resource-Aware Multicore CGRA Architecture for the Edge},   year={2020},  volume={},  number={},  pages={33-39},  doi={10.1109/FPL50879.2020.00017}}

@INPROCEEDINGS{reconfig-on-edge-2,  author={C. {Tan} and A. {Kulkarni} and V. {Venkataramani} and M. {Karunaratne} and T. {Mitra} and L. {Peh}},  booktitle={2016 International Conference on Compliers, Architectures, and Sythesis of Embedded Systems (CASES)},   title={LOCUS: Low-power customizable many-core architecture for wearables},   year={2016},  volume={},  number={},  pages={1-10},  doi={10.1145/2968455.2968506}}

@INPROCEEDINGS{reconfig-on-edge-3,  author={C. {Tan} and M. {Karunaratne} and T. {Mitra} and L. {Peh}},  booktitle={2018 ACM/IEEE 45th Annual International Symposium on Computer Architecture (ISCA)},   title={Stitch: Fusible Heterogeneous Accelerators Enmeshed with Many-Core Architecture for Wearables},   year={2018},  volume={},  number={},  pages={575-587},  doi={10.1109/ISCA.2018.00054}}
\end{document}